\documentstyle[aps]{revtex}

\begin{document}
\title{Quantum data hiding with spontaneous parameter down-conversion}
\author{Guang-Can Guo, Guo-Ping Guo\thanks{%
Electronic address: harryguo@mail.ustc.edu.cn }}
\address{Key Laboratory of Quantum Information, University of Science and Technology\\
of China, Chinese Academy of Science, Hefei, Anhui, P. P. China, 230026}
\maketitle

\begin{abstract}
Here we analyze the practical implication of the existing quantum data
hiding protocol with Bell states produced with optical downconverter. We
show that the uncertainty for the producing of the Bell states with
spontaneous parameter down-conversion should be taken into account, because
it will cause serious trouble to the hider encoding procedure. A set of
extended Bell states and a generalized Bell states analyzer are proposed to
describe and analyze the possible states of two photons distributing in two
paths. Then we present a method to integrate the above uncertainty of Bell
states preparation into the dating hiding procedure, when we encode the
secret with the set of extended Bell states. These modifications greatly
simplify the hider's encoding operations, and thus paves the way for the
implementation of quantum data hiding with present-day quantum optics.

PACS number(s): 03.67.Hk, 03.65.Ud, 89.70.+c
\end{abstract}

\baselineskip12ptWith the development of quantum information theory, more
and more fantastic applications have been explored. It is well known that
quantum mechanics can keep classical and quantum bits secret in a number of
different circumstances. In some scenarios, the bits are kept secret from
eavesdropper while in others, they are kept secret from participants
themselves.

Quantum key distribution\cite{bb84,b92,ekert,or,guo,h1} is the first such
example, which keeps messages secret from any eavesdropper accessing the
output of the quantum channel. As the quantum generalization of the one-time
pad, it is also known as private quantum channel and is most near practical
application. In this case, two parties make use of shared random bits to
create a secure quantum channel between them. Then they can safely transmit
messages with this secure quantum channel. A second example is quantum
secret sharing\cite{h3,guo1}, which aims to share a secret, in the form of
classical or quantum bits, among many parties. Only certain prescribed
combinations of the parties, known as authorized sets, are capable of fully
reconstructing the secret with assistance of local operations and classical
communications. Nothing at all can any unauthorized combination learns about
the secret, even although they can act jointly on their shares or have
quantum communications. The third example is the novel quantum data hiding
recently proposed by Terhal and her cooperators\cite{hb,hb1,hb2}, which
discusses a different security problem in the quantum information field and
explores another new application.

Although quantum data hiding also aims to share a secret between $bi$- or
multi-party, it imposes a much stronger security criterion than quantum
secret sharing. In the quantum data hiding protocols, quantum communications
or channels are prerequisite, even for authorized sets, to revealing the
secret. In Terhal's original protocol of hiding classical bits, $n$ pairs of
Bell states are shared between two parties, Alice and Bob. For each Bell
state, the first qubit goes to Alice and the second to Bob. The secret is
encoded in the number of the state $\left| \Psi \right\rangle ^{-}$ among
those $n$ pairs of Bell states, whose even numbers represent $0$ and odd
numbers denote $1$. The substantial information, which the two sharers could
get about the secret through any sequence of local quantum operations
supplemented by unlimited two-way classical communication (LOCC), is
exponentially small in $n$, the number of Bell states used for the encoding.
Afterward, generalized schemes for hiding classical data in multi-partite
quantum states and hiding quantum data have also been proposed\cite{hb1,hb2}%
. Furthermore, two significant conclusions have been gotten, which provide
the basic descriptions for the problem of quantum data hiding: perfect
quantum data hiding is impossible; the quantum data hiding with pure states
is impossible. In addition, Terhal {\it et al.} discussed the implementation
of the Bell states quantum data hiding protocol by virtue of current quantum
optics setup such as optical down-converter.

Here, we particularly analyze the experimental implications of this Bell
states quantum data hiding protocol with optical down-converter. We show
that the uncertainty for the producing of the Bell states with spontaneous
parameter down-conversion should be taken into account, because it will
cause serious trouble to the hider encoding procedure. Subsequently, we
propose a set of extended Bell states and a generalized Bell states analyzer
to describe and analyze the possible states of two photons distributing
along two paths. Then we present a method to elegantly integrate the above
uncertainty of Bell states preparation into the data hiding procedure, and
encode the secret in a set of the extended Bell states. Compared with the
rigorous security proof for the original quantum data hiding protocol with
Bell states, this modified quantum data hiding protocol can be
straightforwardly argued to maintain similar security. It paves the road for
the experimental implementation of the quantum data hiding with current
quantum optics.

In Terhal's original quantum data hiding scheme\cite{hb}, they proposed to
hide bits in a series of Bell states produced with optical down-converter.
The hider is assumed to have a supply of each of the four Bell states. When
the one-bit secret $b=1,$ the hider picks at random a set of $n$ Bell states
with uniform probability, except that the number of singlets $\left| \Psi
\right\rangle ^{-}$ must be odd. The $b=0$ protocol is the same, except that
the number of singlets must be even. It is well known that the state
produced with the parameter down-conversion is not a Bell state, but a
superposition of the vacuum, a two-photon Bell state, a four-photon state,
etc. In fact this state can be generally written as (unnormalized) 
\begin{equation}
\left| \Sigma \right\rangle =(1+p^{1/2}a_{ij}^{+}+\frac{(p^{1/2}a_{ij}^{+})^2%
}2+o(p))\left| vac\right\rangle
\end{equation}
where $p$ is the probability of producing a pair of Bell state $\left| \Psi
\right\rangle _{ij}^{-}=a_{ij}^{+}\left| vac\right\rangle =\frac 1{\sqrt{2}}%
(h_i^{\dagger }v_j^{\dagger }-v_i^{\dagger }h_j^{\dagger })\left|
vac\right\rangle $, with $h$ and $v$ being the two polarization mode
operators of photon. $o(p)$ represents the terms to produce more
down-conversion photons whose probabilities are smaller than $p^2$ and $%
\left| vac\right\rangle $ is the vacuum state of the down-conversion
photons. Obviously, the hider cannot exactly ascertain when the
down-converter produce photons and whether these photons are in then Bell
state $\left| \Psi \right\rangle ^{-}$. As introducing of postselection
measurements will make quantum data hiding meaningless, this uncertainty
will cause serious problem for the encoding of the quantum data hiding
scheme. It will be very difficult for the hider to pick out $n$ pairs of
Bell states and to ensure that there are exactly even or odd number of
singlets among these states. Although a device of quantum non-demolition
(QND) measurement for Bell states\cite{guo2} can resolve this problem, the
requirement for the uncommon individual photons CNOT gates or single-photon
sources\cite{h8} renders it beyond the reach of the present experimental
conditions.

To cope with this uncertainty in the generation of Bell states, we can
modify the above quantum data hiding protocol in the following way. Consider
an experimental optics setup as shown in Fig. 1. Generally, a pulse of
ultraviolet (UV) light passing through a nonlinear crystal creates a pair of
entangled photons in paths $1$ and $2$. After retroflection, the ultraviolet
pulse creates another pair of photons in paths $3$ and $4$ during its second
passage through the crystal. In view of the uncertainty for the parameter
down-conversion, the total state of photons in paths $1$, $2$, $3$ and $4$
could be written as the following form (unnormalized) 
\begin{eqnarray}
\left| \Xi \right\rangle &=&(1+p^{1/2}a_{12}^{+}+\frac{(p^{1/2}a_{12}^{+})^2}%
2+o(p))\otimes (1+p^{1/2}a_{34}^{+}+\frac{(p^{1/2}a_{34}^{+})^2}2%
+o(p))\left| vac\right\rangle \\
&=&(1+p^{1/2}(a_{12}^{+}+a_{34}^{+})+p(a_{12}^{+}a_{34}^{+}+\frac{%
(a_{12}^{+})^2}2+\frac{(a_{34}^{+})^2}2)+o(p))\left| vac\right\rangle , 
\nonumber
\end{eqnarray}
where $a_{ij}^{+}=$ $\frac 1{\sqrt{2}}(h_i^{\dagger }v_j^{\dagger
}-v_i^{\dagger }h_j^{\dagger })$ is the creation operator for the singlet
state $\left| \Psi \right\rangle ^{-}$, and $\left| vac\right\rangle $ is
the vacuum state of the four paths. Obviously we have a probability of the
order of $p^2$ to have totally four photons in the four paths $1$, $2$, $3$
and $4$, which are in the state (unnormalized): 
\begin{equation}
\left| \Theta \right\rangle =(a_{12}^{+}a_{34}^{+}+\frac{(a_{12}^{+})^2}2+%
\frac{(a_{34}^{+})^2}2)\left| vac\right\rangle .
\end{equation}
This state can be also written as (unnormalized) 
\begin{eqnarray}
\left| \Theta \right\rangle &=&\left| \Phi \right\rangle _{13}^{+}\left|
\Phi \right\rangle _{24}^{+}-\left| \Phi \right\rangle _{13}^{-}\left| \Phi
\right\rangle _{24}^{-}-\left| \Psi \right\rangle _{13}^{+}\left| \Psi
\right\rangle _{24}^{+}+\left| \Psi \right\rangle _{13}^{-}\left| \Psi
\right\rangle _{24}^{-}  \nonumber \\
&&+\left| \Gamma \right\rangle _{13}^{+}\left| \Upsilon \right\rangle
_{24}^{+}+\left| \Gamma \right\rangle _{13}^{-}\left| \Upsilon \right\rangle
_{24}^{-}+\left| \Upsilon \right\rangle _{13}^{+}\left| \Gamma \right\rangle
_{24}^{+}+\left| \Upsilon \right\rangle _{13}^{-}\left| \Gamma \right\rangle
_{24}^{-}-\left| \Omega \right\rangle _{13}^{+}\left| \Omega \right\rangle
_{24}^{+}-\left| \Omega \right\rangle _{13}^{-}\left| \Omega \right\rangle
_{24}^{-}.
\end{eqnarray}
Here $\left| \Phi \right\rangle _{ij}^{\pm }=\frac 1{\sqrt{2}}(h_i^{\dagger
}h_j^{\dagger }\pm v_i^{\dagger }v_j^{\dagger })\left| vac\right\rangle $
and $\left| \Psi \right\rangle _{ij}^{\pm }=\frac 1{\sqrt{2}}(h_i^{\dagger
}v_j^{\dagger }\pm v_i^{\dagger }h_j^{\dagger })\left| vac\right\rangle $
are the common four Bell states, which constitute a set of complete bases in
the Hilbert space $H1$. This space represents the case that there is one and
only one photon in each of two paths $i$ and $j$. The states $\left| \Gamma
\right\rangle _{ij}^{\pm }=\frac 12(h_i^{\dagger }h_i^{\dagger }\pm
v_j^{\dagger }v_j^{\dagger })\left| vac\right\rangle ,$ $\left| \Upsilon
\right\rangle _{ij}^{\pm }=\frac 12(v_i^{\dagger }v_i^{\dagger }\pm
h_j^{\dagger }h_j^{\dagger })\left| vac\right\rangle $ and $\left| \Omega
\right\rangle _{ij}^{\pm }=\frac 1{\sqrt{2}}(h_i^{\dagger }v_i^{\dagger }\pm
h_j^{\dagger }v_j^{\dagger })\left| vac\right\rangle $ correspond to the
case that there are two photons concentrating in one path and with no photon
in the other path. These six states can also be regarded as a set of
complete generalized Bell bases in the Hilbert space $H2$, where two photons
concentrate in one certain path. Thus there are generally ten Bell-type
states involving two photons and two paths, which respectively belong to two
sets of bases. Obviously those two sets of bases lie in two different
Hilbert spaces, $H1$ and $H2$.

In the first step of the present modified quantum data hiding protocol, the
hider measures the photons from the paths $1$ and $3$ with an optical setup
as shown in Fig. 1\cite{guo4,guo3}. When there are coincidence clicks
between two same polarization mode detectors $D_V^u$ and $D_V^d$ (or $D_H^u$
and $D_H^d$), the two photons in paths $1$ and $3$ are measured in either
the state $\left| \Phi \right\rangle _{13}^{+}$ or the state $\left| \Omega
\right\rangle _{13}^{+}$. And then the two photons in paths $2$ and $4$ are
obviously collapsed into the state $\left| \Phi \right\rangle _{24}^{+}$ or
the state $\left| \Omega \right\rangle _{24}^{+}$. Similarly, when there are
coincidence clicks between two different polarization mode detectors $D_H^u$
and $D_V^d$ (or $D_V^u$ and $D_H^d$), two photons in paths $1$ and $3$ are
measured in either the state $\left| \Phi \right\rangle _{13}^{-}$ or the
state $\left| \Omega \right\rangle _{13}^{-}$. And thus the two photons in
paths $2$ and $4$ are collapsed into the state $\left| \Phi \right\rangle
_{24}^{-}$ or the state $\left| \Omega \right\rangle _{24}^{-}$. Analogous
to the existing Bell states analyzer with linear optics, this optical setup
as shown in Fig.1 can be regarded as a general Bell analyzer (GBA). The GBA
can divide the ten general Bell states into three classes: $\left| \Phi
\right\rangle _{ij}^{+}$ and $\left| \Omega \right\rangle _{ij}^{+}$ as the
first class, $\left| \Phi \right\rangle _{ij}^{-}$ and $\left| \Omega
\right\rangle _{ij}^{-}$ as the second class, and the others as the third
class.

According to the measurement results of the photons in path $1$ and $3$, the
hider can conveniently pick out $n$ pairs of photons in paths $2$ and $4$,
which are randomly in the above three classes general Bell states. When the
one-bit secret $b=1$, the hider picks out odd number of the first class
states (can be either $\left| \Phi \right\rangle _{ij}^{+}$ or $\left|
\Omega \right\rangle _{ij}^{+}$) among these $n$ pairs of states chosen at
random. For the case $b=0$, the hider chooses even number of the first class
states in those $n$ pairs of general Bell states. This encoding procedure is
straightforward and effortless. The uncertainty caused by the parameter down
conversion is ingeniously integrated into the encoding states.

To hide the secret $b$, the $n$ pairs of photons in paths $2$ and $4$ are
sent to the sharers, with the photons in path $2$ to Alice and path $4$ to
Bob respectively. To completely decode the secret, a quantum channel between
Alice and Bob is opened up and one sharer's photons, say Alice, are sent to
the other, as Bob. Then Bob can measure these photons with the same general
Bell states analyzer (GBA) as the hider has used. Simply count the number of
the first class states measured (the number of the coincidence clicks
between two same mode detectors), the sharers can easily figure out the
parity and then the secret.

The rigorous proof for the security of the present quantum data hiding
protocol with ten generalized Bell states is involuted and will be presented
in other place\cite{guo5}. Here we propose a simple but suggestive argument,
which states that the present modified quantum data hiding protocol can be
at least $2/5$ times as security as Terhal's original scheme.

The secret $b$ is encoded in the parity of the total number of the states $%
\left| \Phi \right\rangle ^{+}$ and $\left| \Omega \right\rangle ^{+}$ in
the tensor product of $n$ general Bell states of the above two sets. We can
then assume that among these $n$ pairs of encoded states, there are $m$
pairs of states of the set $S1=\{\left| \Phi \right\rangle ^{\pm },\left|
\Psi \right\rangle ^{\pm }\}$ and $(n-m)$ pairs of states of the set $%
S2=\{\left| \Gamma \right\rangle ^{\pm },\left| \Upsilon \right\rangle ^{\pm
},\left| \Omega \right\rangle ^{\pm }\}$. The security analyze for quantum
data hiding is equal to bounding the mutual information $I(b:M)$ the sharers
can get about the secret $b$ with LOCC operations $M$. Generally, there are
two manners for the sharers to decode the secret $b$. In the first method,
the two sharers do not try to separate the two sets of states, and directly
act on the tensor product state of all these $n$ pairs of states. Any
sequence of LOCC operations is allowed for the sharers. In the second
method, the two sharers firstly divide those $n$ pairs of states into two
sets $S1$ and $S2$ with some LOCC operations. Afterward, they separately
decode the number $n_1$of the state $\left| \Phi \right\rangle ^{+}$ from
the $m$ pairs of $S1-$set states and the number $n_2$ of the state $\left|
\Omega \right\rangle ^{+}$ from the $(n-m)$ pairs of $S2-$set states. By
combining the parity $b_1$of the number $n_1$and the parity $b_2$ of the
number $n_2$, the two sharers can learn the secret $b=b_1\oplus b_2$, with $%
\oplus $ being the addition modulo two.

As the sharers can do any sequence of LOCC operations in decode procedure,
the second method is in fact a particular example contained in the first
general method. Obviously, the states of the two sets $S1$ and $S2$ lie in
two different Hilbert space $H1$ and $H2,$ and respectively represent the
case the two photons distribute in two paths or concentrate in one path.
Thus we argue that the sharers cannot loss any advantage for decoding by
firstly separate the two sets of different Hilbert space states. The mutual
information $I(b:M)$ the two sharers getting about the secret $b$ with the
second particular method will not be less than that by through the first
general method. We can then prove the security of the present quantum data
hiding protocol by analyze the second particular decoding method.

Since the sharers can theoretically do any sequence of LOCC operations on
the photons, Alice and Bob can easily separate the states of the two sets $S1
$ and $S2$ with some quantum non-demolition devices as
photon-Fock-state-filter. Then the two sharers separately decode the parity $%
b1$ and $b2$ from the $n$ pairs of $S1-$set states and the $(n-m)$ pair of $%
S2-$set states. It can be proved that the sharers can exactly decode the
parity $b2$ from the $(n-m)$ pair of $S2-$set states. With the result from
the original quantum data hiding protocol with $S1-$set Bell states, the
mutual information $I(b1:M)$\cite{h14} the sharers can get about the parity $%
b1$ with LOCC operation is bounded by $\delta H(b1)$\cite{hb}, where $\delta
=1/2^{m-1}$ and $H(B1)$ is the Shannon information of the hidden bit. Thus
the mutual information $I(b:M)$ the sharers can get about the secret $%
b=b1\oplus b2$ with the second method by separately acting on the two sets
is only bounded by $\delta H(b1)=H(B1)/2^{m-1}$.

We have argued this mutual information getting from the second method is
also the bound of that the two sharers can get with any sequence of LOCC
operations. It is easy to see that the two photons in path 2 and 4 has a
probability of $2/5$ to be prepared in the $S1-$set states in the present
quantum data hiding scheme with spontaneous parameter down-conversion. Thus,
to achieve the same level of security, the present protocol needs $5/2$
times as many pairs of states as the original quantum data hiding with $S1-$%
set Bell states.

In conclusion, we have analyzed the practical implication of the existing
quantum data hiding protocol with Bell states produced with optical
down-converter. We showed that the uncertainty for the producing of the Bell
states with spontaneous parameter down-conversion should be taken into
account, because it will cause serious trouble to the hider encoding
procedure. A set of extended Bell states and a generalized Bell states
analyzer are proposed to describe and analyze the possible states of two
photons distributing in the two paths. Then we presented a method to
integrate the above uncertainty of Bell states preparation into the dating
hiding procedure, when we encode the secret with a set of extended Bell
states. These modifications greatly simplify the hider's encoding
operations. With the result from the origin protocol, the present modified
quantum data hiding scheme is argued to have similar security. It paves the
road for the experimental implementation of the quantum data hiding with
present-day quantum optics.

We thank Barbara Terhal for the discuss in the security proof of the present
protocol. This work was funded by National Fundamental Research
Program(2001CB309300), National Natural Science Foundation of China, the
Innovation funds from Chinese Academy of Sciences, and also by the
outstanding Ph. D thesis award and the CAS's talented scientist award
entitled to Luming Duan.

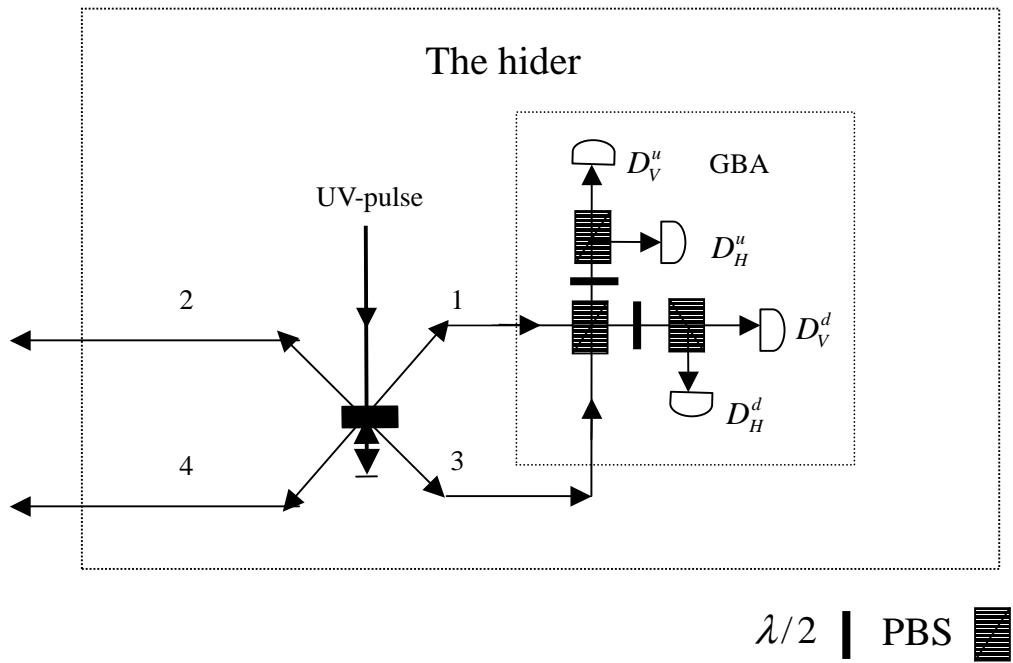
Figure1: The schematic setup for the modified quantum data hiding protocol
with extended Bell states. A pulse of ultraviolet (UV) light passing through
a nonlinear crystal creates the ancillary pair of entangled photons in paths 
$1$ and $2$. After retroflection during its second passage through the
crystal, the ultraviolet pulse can create another pair of photons in paths $%
3 $ and $4$. Then there is a probability of order of $p^2$ to have four
photons in the four paths $1$, $2$, $3$ and $4$. The $\lambda /2$ plates are
used to implement Hardmard operations, which transform $h$ mode photon into $%
h-v,$ and $v$ mode into $h+v$. To encode secret, the hider measures the
photons from the path $1$ and $3$ with an general Bell states analyzer(GBA),
and picks out $n$ pairs of photons in paths $2$ and $4$, which are
respectively sent to the two sharers, Alice and Bob. In the secret decoding
procedure, Alice and Bob cooperatively measure the photons from paths $2$
and $4$ with the same analyzer(GBA).


\begin{references}
\bibitem{bb84}  C. H. Bennett, and G. Brassard, Advances in Cryptology:
Proceedings of Ctypto84, August 1984, Springer-Verlag, p.475.

\bibitem{b92}  C. H. Bennett, Phys. Rev. Lett. 68, 3132 (1992).

\bibitem{ekert}  A. Ekert, Phy. Rev. Lett. 67, 661 (1991).

\bibitem{or}  L. Goldenberg, and L. Vaidman, Phys. Rev. Lett. 75, 1239
(1995).

\bibitem{guo}  G. P. Guo, C. F. Li, B. S. Shi, J. Li, and G. C. Guo, Phys.
Rev. A 64, 042301 (2001).

\bibitem{h1}  A. Ambainis, M. Mosca, A. Tapp, and R. de Wolf. In IEEE
Symposium in Foundations of Computer Science (FOCS), 547 (2000).

\bibitem{h3}  R. Cleve, D. Gottesman, and H. K. Lo. Phys. Rev. Lett. 83, 648
(1999).

\bibitem{guo1}  G. P. Guo, and G. C. Guo Phys. Lett. A 310, 247 (2003).

\bibitem{hb}  B. M. Terhal, D. P. Divincenzo, and D. W. Leung, Phys. Rev.
Lett. 86, 5807 (2001).

\bibitem{hb1}  D. P. Divincenzo, P. Hayden, and B. M. Terhal.
quant-ph/0207147.

\bibitem{hb2}  D. P. Divincenzo, D. W. Leung, and B. M. Terhal.
quant-ph/0103098.

\bibitem{guo2}  G. P. Guo, C. F. Li, and G. C. Guo, Phys. Lett. A 286,401
(2001).

\bibitem{h8}  E. Knill, R. Laflamme, and G. Milburn, Nature 409, 46 (2001).

\bibitem{guo4}  G. P. Guo, and G. C. Guo quant-ph/0208071.

\bibitem{guo3}  G. P. Guo, and G. C. Guo quant-ph/0301009.

\bibitem{guo5}  G. P. Guo, and G. C. Guo in preparation.

\bibitem{h14}  T. M. Cover and J. A. Thomas, Elements of Information Theory
(Wiley, New York, 1991).

{\bf Figure Captions:}
\end{references}
\end{document}